\documentclass[10pt,a4paper,twocolumn]{article}
\usepackage[top=18truemm,bottom=18truemm,left=16truemm,right=10truemm]{geometry}

\usepackage{cite}
\usepackage{amsmath,amssymb,amsfonts}
\usepackage{algorithmic}
\usepackage{graphicx}
\usepackage{textcomp}
\usepackage{caption}
\usepackage{subfigure}
\usepackage{url}

\title{Design of Ad Hoc Wireless Mesh Networks \\ Formed by Unmanned Aerial Vehicles \\ with Advanced Mechanical Automation}

\author{Ryoichi Shinkuma\\
Graduate School of Informatics,Kyoto University\\
Yoshidahon-machi, Sakyo-ku, Kyoto, Japan, 606-8501, Japan,\\
E-mail: shinkuma@i.kyoto-u.ac.jp \\ 
and \\
Narayan B. Mandayam\\
WINLAB (Wireless Information Network Laboratory), Rutgers University\\
Technology Centre of New Jersey, 671 Route 1 South, North Brunswick, NJ, 08902-3390, USA}

\begin{document}

\maketitle

\section*{Abstract}
{Ad hoc} wireless mesh networks formed by unmanned aerial vehicles (UAVs) equipped with wireless transceivers (access points (APs)) are increasingly being touted as being able to provide {a} flexible ``on-the-fly'' communications infrastructure that can collect and transmit sensor data from sensors in remote, {wilderness,} or disaster-hit areas. 
Recent advances in {the} mechanical automation of UAVs {have} resulted in separable APs and replaceable batteries that can be carried by UAVs and placed at arbitrary locations in the field.  
{These} advanced mechanized UAV mesh networks pose interesting questions in terms of {the} design of {the} network architecture {and the} optimal UAV scheduling algorithms. 
This paper studies a range of network architectures {that depend} on the mechanized automation (AP separation and battery replacement) capabilities of {UAVs} and proposes heuristic UAV scheduling algorithms for each network architecture, which are {benchmarked} against optimal designs.

\noindent{\bf unmanned aerial vehicle, wireless mesh network, battery replacement}


\section{Introduction}\label{sec:introduction}

{Micro} or small unmanned aerial vehicles (UAVs) have been receiving increasing attention in military, commercial, and social applications \cite{Bekmezci2013}.
{These} UAVs are expected to be emerging solutions for surveying areas {in which} humans and ground vehicles cannot easily {enter, such as} untouched {wilderness areas and} disaster-damaged areas \cite{Andre2014,Erdelj2016,Ahmed2008,Saggiani,Abdelkader2013}. 
{They} are also operated as an {ad hoc} wireless mesh network infrastructure that connects such isolated areas to a communication infrastructure \cite{Felice2014,Daniel2009,Quaristch2010}, which could be a promising solution{, particularly} for collecting sensor data obtained in {these} areas.
{The benefits of airborne relaying, in which a UAV provides an interconnection between an isolated area and a communication infrastructure, have been discussed \cite{Feng2007}.}
It was reported {that} airborne connection provides better connectivity and throughput than ground {connections} because {three-dimensional} positioning of relaying nodes {provides} line-of-site (LOS) propagation and {suppresses} shadowing and fading effects more effectively. 
Multihop airborne relaying {and} aerial wireless mesh networks {that use} a multi UAV system {have also been} proposed \cite{Felice2014,UAVNet2012}. 
Although those networks cover longer {distances} than single-hop {networks}, {the} technical problems of routing and scheduling {are} more complicated. 
Researchers have been working on routing protocols and algorithms and battery recharging scheduling {for maintaining} the connectivity of wireless mesh networks. 

\begin{figure}[t]
\begin{center}
\includegraphics[width=0.7\linewidth]{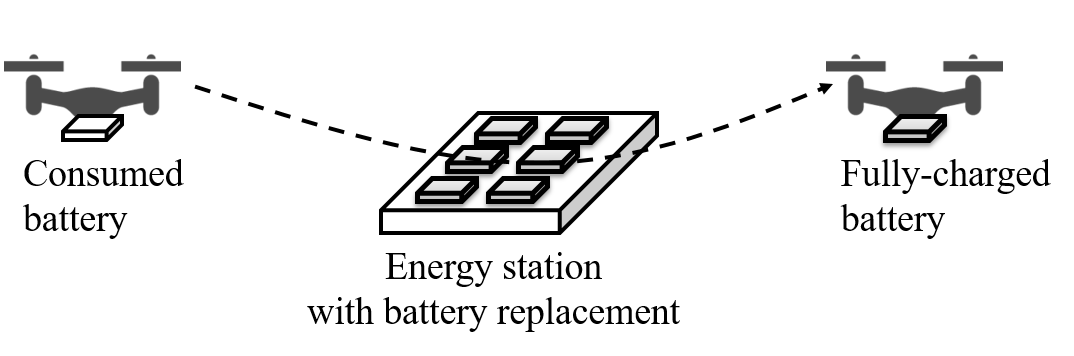}
\caption{Automatic battery replacement}
\label{fig:replace}
\end{center}
\end{figure}

\begin{figure}[t]
\begin{center}
\includegraphics[width=0.7\linewidth]{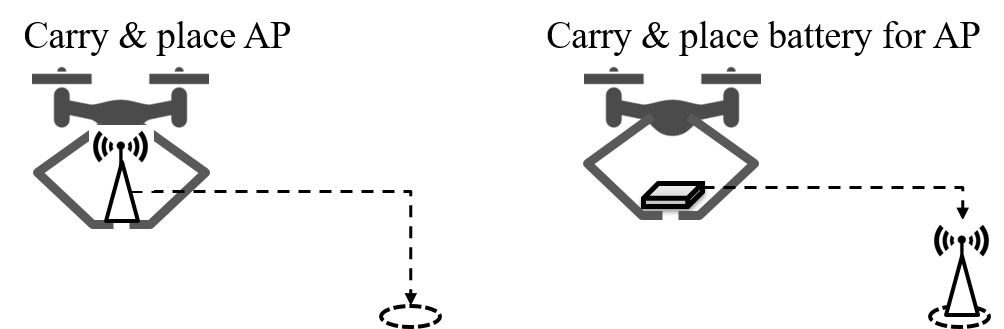}
\caption{AP and battery carrying and placement{,} enabled by load manipulation}
\label{fig:load}
\end{center}
\end{figure}

However, networks built on the basis of {conventional} techniques are not {sufficiently} effective in terms of sustainability because earlier works have not considered the following {assumptions}:
(1) {Wireless} access points (APs) can be separable from UAVs and carried by UAVs{. Therefore,} UAVs do not need to keep flying once APs are placed at {appropriate} positions for connectivity{.}
(2) UAV and AP batteries can be {replaced} and carried by UAVs. UAVs do not need to wait at the {energy station} until their batteries are fully charged.
The recent development of mechanical automation for UAVs has been making these assumptions realistic.
First, the automatic battery-replacement technology, which is also called battery swapping, for UAVs has been recently well-discussed \cite{Swieringa2010, Kemper F2011, Suzuki2012, Toksoz2011, Ure2015, Fujii2013}.
As illustrated in Figure~\ref{fig:replace}, the automatic battery-replacement technology {enables} UAVs to replace their {discharged} batteries {with} new ones and start flying immediately{,} without waiting until they are fully charged at the energy station.
{Moreover}, the technology of load manipulation by UAVs has been developed \cite{Pounds2011, Palunko2012, Barteld2016, ieee-aerialrobotics}.
With this technology, as illustrated in Figure~\ref{fig:load}, UAVs can carry small APs and batteries for the APs and place them at {appropriate} positions.

In this paper, we propose a new design {for} wireless mesh networks formed by UAVs under the assumption that {both} batteries and APs are replaceable and separable from UAVs and are carried and placed at the {appropriate} positions by the mechanical automation of UAVs. 
We {present} possible design models of UAV-formed mesh networks and discuss the {advantages} and {disadvantages} of each model. 
We also consider the number of UAVs {required} for maintaining the connectivity as a primary metric of how feasible the design model is and show numerical results {obtained} through computer simulations.
%
{In addition, we} examine and present the number of required batteries and the throughput performance {of} our system.

Our contributions {are} summarized as {follows:} 
1) we present a novel approach that addresses mesh networks formed by UAVs with replaceable batteries and separable APs; 
2) we {present} possible design options of UAV-formed mesh networks and compare them;
{and} 3) we study the feasibility of designing such mechanically automated networks in terms of the {numbers} of required UAVs and batteries and {the} throughput, which are evaluated {by computer simulations.}
The differences between this paper and our previous paper \cite{UEMCON2016} {are} summarized as {follows:}
1) this paper has developed the mathematical formulation of our system to discuss the baseline and the lower-bound performances in the simulation results;
2) the previous paper assumed a linear battery model without considering the non-linear {characteristics} of the battery;
3) the previous paper assumed an unlimited number of batteries{,} while this paper evaluates the required number of batteries;
and 4) {the} throughput performance of UAV-formed mesh networks was not evaluated in the previous paper.

The remainder of this paper is organized as follows{:}
Section \ref{sec:related} discusses related work.
Section \ref{II} presents the model, the problem statement, and the algorithm for UAV operation of the proposed system.
{Then,} Section \ref{sec:feasibility} {demonstrates} the model and {presents} the {results} of the feasibility evaluation{, which uses} a realistic battery model.
{Next,} the throughput performance analysis {is presented} in Section \ref{sec:throughput}.
Finally, Section \ref{VII} concludes this paper.

\section{Related work}
\label{sec:related}
This section discusses the prior works related to this paper.
First{, we discuss} the automatic battery-replacement technology, which is one of the basic requirements of UAVs in our system.
A load-manipulation technology for UAVs is assumed in our system since {UAVs} need to carry APs {and place them} at {predetermined} positions and carry batteries for APs.
Finally, we discuss prior works that have addressed scheduling problems in UAV systems.

\subsection{Battery replacement}
{In 2010,} Swieringa et al. demonstrated a battery swapping mechanism and online algorithms {for addressing} resource management, vehicle health monitoring, and precision landing onto the battery swapping mechanism's landing platform \cite{Swieringa2010}.

There were many research efforts on battery replacement technology in 2011.
Kemper et al. proposed three station designs for refilling platforms and one concept for battery exchange platforms \cite{Kemper F2011}. They also analyzed the economic feasibility of automatic consumable replenishment stations, considered two types of {stations} (container refilling and container exchange), and discussed the application of these systems. They {asserted that} refilling platforms better suit {low-coverage} unmanned aerial systems {(UASs),} while exchange stations allow high coverage with fewer UAVs.
In another work, they compared different solutions for various modules of an automated battery replacement system for UAVs \cite{Suzuki2012}. They also proposed a ground station capable of swapping a UAV's batteries and discussed prototype components and tests of some of the prototype modules. They concluded {that} their platform is well-suited for high-coverage requirements and is capable of handling a heterogeneous UAV fleet.

Toksoz et al. introduced a hardware platform for automated battery changing and charging for multiple UAV agents \cite{Toksoz2011}. {From the results of experiments in an indoor flight test facility, they concluded that their change/charge station has sufficient capability and robustness in the context} of a multi-agent, persistent mission where surveillance is continuously required over a specified region.
In 2015, they presented the development and hardware implementation of an autonomous battery maintenance mechatronic system that significantly extends the operational time of {battery-powered} small-scale UAVs \cite{Ure2015}. {This} automated system quickly swaps a depleted battery of a UAV with a replenished one while simultaneously recharging several other batteries.

In 2013, Fujii et al. proposed the concept of ``Endless Flyer'': they developed an automatic battery replacement mechanism that allows UAVs to fly continuously without manual battery replacement and suggested scalable and robust {applications} for the system. They conducted an initial experiment using this system and successfully assessed the possibility of continuous surveillance in both indoor and outdoor environments \cite{Fujii2013}.

\subsection{Load manipulation}
In 2011, Pounds et al. analyzed key challenges encountered when lifting a grasped object and transitioning into laden free-flight \cite{Pounds2011}. They determined stability bounds in which the changing mass-inertia parameters of the system due to the grasped object will not destabilize the flight controller.  They demonstrated grasping and retrieval of a variety of objects while hovering, without touching the ground, using the Yale Aerial Manipulator testbed.

In 2012, Palunko et al. tackled the challenging problem of using quadrotors to transport and manipulate loads safely and efficiently. Aerial manipulation is extremely important in emergency rescue missions {and} in military and industrial applications \cite{Palunko2012}. They described and summarized two possible approaches that enable agile and safe load transportation using a single quadrotor UAV: 1) an adaptive controller {that considers} changes in the center of gravity and 2) {an} optimal trajectory {generator} based on dynamic programming for swing-free maneuvering. From the simulation results, they verified the validity of the proposed algorithms. They also presented experimental results of the proposed optimal swing-free trajectory tracking {approach}.

In a project named AEROWORKS, Bartelds et al. proposed a solution that combines control of the aerial manipulator's end-effector position with an innovative design approach of aerial manipulation systems consisting of both active and passive joints \cite{Barteld2016}. The approach aimed at limiting the influence of impacts on the controlled attitude dynamics {to} allow the aerial manipulator to remain stable during and after impact. The experimental results showed that the proposed approach and the developed mechanical system achieve stable impact absorption without bouncing away from the interacting environment.

\subsection{Scheduling}
Cummings and Mitchell {performed} {pioneering} work {on} scheduling in UAV systems \cite{Cummings2007}: {To} study how levels of automation affect UAV knowledge-based {missions} and payload management control {loops} from a human supervisory control perspective, {they conducted} a simplified simulation of multiple UAVs operating independently of one another. The goal was to determine how increasing levels of automation affected operator performance {to} identify possible future automation strategies for multiple UAV scheduling.

{In 2013}, Kim et al. developed a mixed-integer linear program (MILP) model {for formalizing} the problem of scheduling a system of UAVs and multiple shared bases in disparate geographic locations \cite{Kim2013}. {In practice, their} approach allowed for a long-term mission to receive uninterrupted UAV service by successively handing off the task to replacement UAVs served by geographically distributed shared bases.

Felice et al. investigated the utilization of low-altitude aerial mesh networks {of} small UAVs (SUAVs) to re-establish connectivity among isolated end-user devices located on the ground \cite{Felice2014}. In particular, they addressed the problem of energy lifetime and proposed a distributed charging scheduling scheme through which {persistent coverage} of SUAV-based mesh nodes can be guaranteed {in an} emergency scenario.
They presented a distributed algorithm {that enables SUAV-based mesh nodes to autonomously decide} when to recharge. Their algorithm was designed on the basis of the following requirements: (i) it attempts to preserve the {connectivity index} by giving precedence to SUAV-based mesh nodes whose departure will not cause the partitioning of the aerial mesh, and (ii) it accounts for the recharging need of each SUAV-based mesh node on the basis of its residual energy.


\section{Proposed System}\label{II}
\subsection{UAV-AP models}
\label{sec:model}
We consider {multiple} representative classes of UAV-AP models{,} as summarized in Table 1.
The first classification {divides the models into UAV-AP joint (JNT) models and UAV-AP separate (SPT) models}.
In the former, UAVs work as APs when they stay at the predetermined AP positions.
In the latter, UAVs and APs work separately as UAVs and APs.
Furthermore, as shown in Table \ref{UAV-AP_models}, 
the second classification is based on the battery replenishment strategies {and classifies the models into charging (CH) and replacement (RP) models}.
Thus{,} the UAV-AP models are JNT-CH, JNT-RP, SPT-CH, and SPT-RP.

\begin{table}[bt]
    \caption{UAV-AP models. JNT and SPT stand for {joint and separate}. CH and RP {stand for charge and replacement.}}
    \centering
    \label{UAV-AP_models}
    \scalebox{0.82}{
 \begin{tabular}{|c|c|c|c|c|c|c|}
 \hline
 & & \multicolumn{2}{|c|}{Battery replenishment } & \multicolumn{2}{|c|}{Battery replenishment} \\
 & & \multicolumn{2}{|c|}{for UAV} & \multicolumn{2}{|c|}{for AP} \\ \cline{3-6}
 Model & \shortstack{UAV-AP\\joint model} & Charged & Battery & Charged & Battery \\ 
 & & by ES & replacement  & by UAV  & replacement \\  \hline 
 JNT-CH & \checkmark & \checkmark &  & & \\ \hline
 JNT-RP & \checkmark & & \checkmark  & & \\ \hline
 SPT-CH & & \checkmark &  & \checkmark & \\ \hline
 SPT-RP & & & \checkmark &  & \checkmark \\ \hline
 \end{tabular} 
 }
\end{table}

\subsection{UAV-AP joint case} \label{sec:joint}
\subsubsection{System model}\label{sec:system_joint}
\begin{figure}[t]
 \begin{center}
  \includegraphics[width=0.9\linewidth]{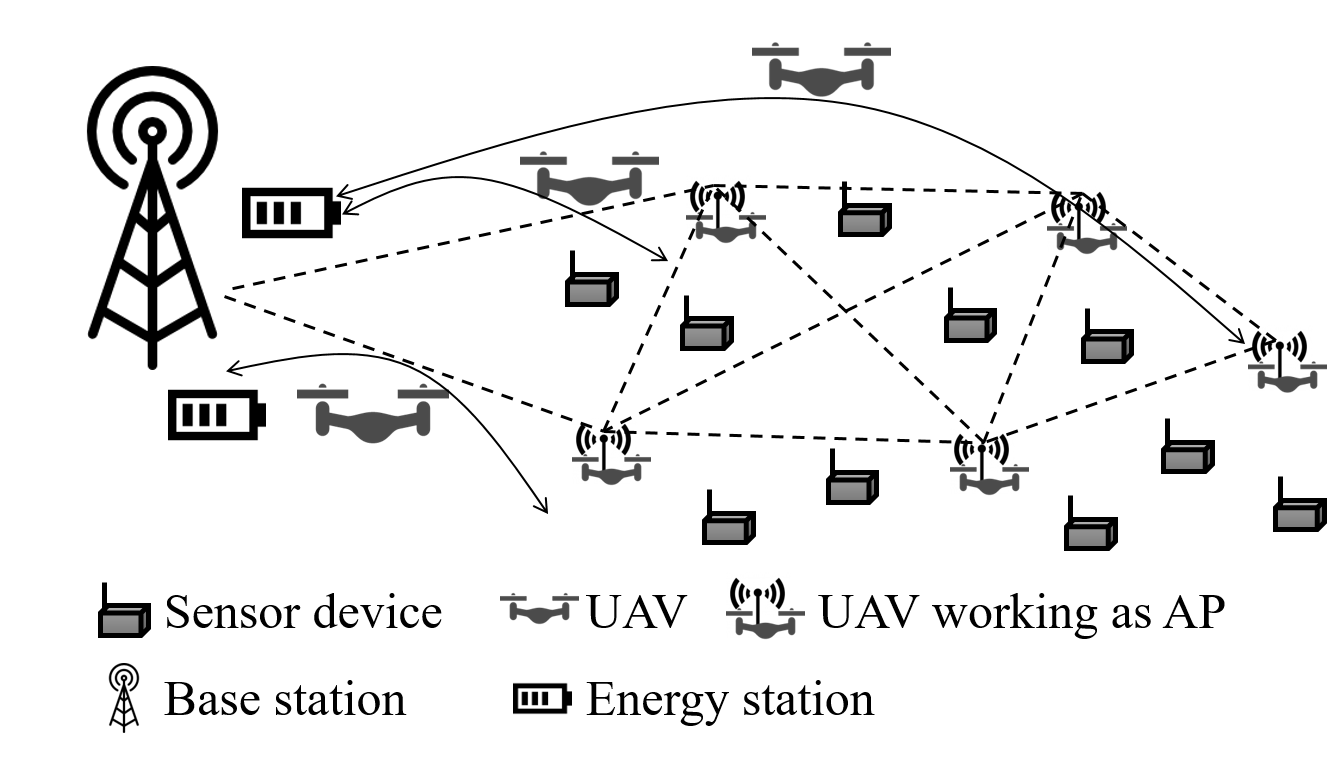}
  \caption{System model in {the} UAV-AP joint case}
  \label{fig:joint_model}
 \end{center}
\end{figure}

The proposed system in the UAV-AP joint case is illustrated in Figure~\ref{fig:joint_model}.
The basic components of the system are a base station (BS), UAVs, and sensor devices.
As illustrated in Figure~\ref{fig:joint_model}, some of the UAVs work as APs at the predetermined AP positions.
The BS is connected to the internet {and forwards} sensor data to users {over} the internet{,} while the UAVs at the AP positions form a wireless mesh network for collecting sensor data from sensor devices and forward {the data} to the BS in a multihop manner.
The BS is operated with {an} energy infrastructure{,} while {the UAVs} are operated with batteries. 
Although sensor devices are also operated with batteries, this paper does not discuss energy issues for sensor devices because, in general, their lifetimes are {sufficiently long,} 
even with small batteries \cite{Tozlu2012}.

JNT-CH and JNT-RP in Table \ref{UAV-AP_models} follow the UAV-AP joint model, in which UAVs work as APs when they stay at the AP positions.
UAVs {not} working as APs are necessary for sustaining the lifetime of the network{.}
{When} a UAV with {a} longer battery lifetime arrives at {a} predetermined position, the UAV working as an AP at {that} position is replaced {with} that one.
The difference between {the JNT-CH and JNT-RP models} is that, while UAVs need to wait at the ES until their batteries are fully charged in JNT-CH, they can replace their batteries with {well-charged ones} in advance at the ES in JNT-RP.
In comparison with the UAV-AP separate models, the UAV-AP joint models have advantages {due to} their simplicity: UAVs {require} no mechanisms {for charging} or {replacing} AP batteries and do not {need} to carry APs{.}
{This} also enables us to {use a simple} scheduling algorithm for UAVs.

\par
A realistic application of our system model in Figure~\ref{fig:joint_model} is research and prevention of disasters in untouched wilderness areas and disaster-damaged areas, where it is often difficult but important to {collect} sensor data.
An example of such {an application} is detection of forest fires\cite{forest-fire}. By collecting sensor data related to {vegetation} in real time, we can quickly detect when a forest fire might occur; the collected data {are} used to estimate {the} hydric stress and risk index, which enables early detection of forest fires.

\subsubsection{Problem formulation}\label{sec:formulation_joint}
In the previous section, we presented {a} system model in the UAV-AP joint case and described the details of the two UAV-AP joint models in Table \ref{UAV-AP_models}.
In general, as the number of UAVs operated in the system increases, it becomes {increasingly} challenging to schedule {them}, place APs, place ESs, and manage the {interactions among} these aspects. 
In the rest of the paper, we work under the assumption that the smaller the number of {UAVs} needed to achieve {specified} system level performance goals {is}, the more desirable {the configuration}.

\begin{figure}[t]
 \begin{center}
  \includegraphics[width=0.7\linewidth]{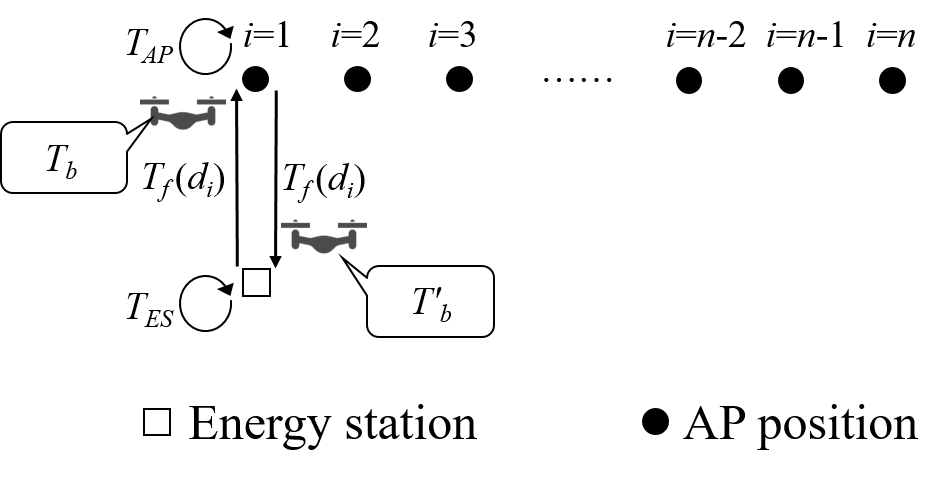}
  \caption{Model for {the} problem formulation}
  \label{fig:theory}
 \end{center}
\end{figure}

To simplify the problem formulation, in this paper, {we assume that} UAVs fly from an AP position to the closest ES or from {an} ES to {an} AP {position; i.e.,} they do not directly fly between AP positions.
Suppose that $i$ and $N${, which are shown in Figure~\ref{fig:theory}, represent} the identification {number} and the number of AP positions and {that $d_i$ denotes} the distance from AP position $i$ to the closest ES.
$T_{ES}$, $T_{AP}$, $T_{f}$, $T_{b}$, and $T'_{b}$ denote the {required time duration} for charging or replacing the battery at {the} ES, the battery lifetime of the AP, the one-way trip time between {the} ES and the AP position, the remaining battery lifetime of {the} UAV from {the} ES to the AP position, and the remaining battery lifetime of {the} UAV from the AP position to {the ES, respectively}.
The problem of minimizing the number of UAVs can be stated as
%
\begin{eqnarray}
\label{eq:objective} 
\min_{\rho} M ( \mbox{D} ),
\end{eqnarray}
\noindent
where $M$ denotes the total number of UAVs required for maintaining the sustainability of the network in the model, $\mbox{D}$ is the set of distances between each AP and the closest ES, {and} $\rho$ is {the} scheduling rule for UAVs.
Every UAV in the system is autonomously operated {according to} $\rho$ in a distributed manner.
When UAVs are operated, three constraints {must} be satisfied:
\begin{eqnarray}
2 T_f(d_i) +T_{ES} & < &T_{AP}~~(\forall i) \label{eq:constraint1}\\
T_f(d_i) &  < & T_b~~(\forall i) \label{eq:constraint2}\\
T_f(d_i) & < & T_b'~~(\forall i) \label{eq:constraint3},
\end{eqnarray}
\noindent
where {constraint} (\ref{eq:constraint1}) means that the AP lifetime must be longer than the {sum} of the roundtrip flying time of {the} UAV between {the} ES and the AP position and the battery charging or {replacement} time for every AP position.
Otherwise, the network could not be sustained {due to the depletion of the} battery of one or more APs.
{Constraint} (\ref{eq:constraint2}) means that the UAV should not {suffer complete battery discharge while} flying to the AP position.
$T_b$ is the remaining battery lifetime after the battery has been charged or replaced at {the} ES.
{Constraint} (\ref{eq:constraint3}) means that the UAV should not {suffer complete battery discharge while} flying back to {the} ES.
$T'_b$ in the UAV-AP joint model is the remaining battery lifetime after the battery has been consumed by {a} UAV that worked as {an} AP.

In the joint AP model, at least one UAV must work as an AP at each predetermined AP position.
Therefore, $M$ should be {equal to} $N$ + $m$, where $N$ is the number of AP positions and $m$ is the number of redundant UAVs not working as APs.
{In a basic operation in this model,} the number of redundant UAVs is one for each AP position; {therefore}, {the baseline required number} of UAVs in this model is $2N$.
The minimum number of required UAVs is $N+1$, which means that only one redundant UAV is operated to maintain {a} network that consists of $N$ AP positions.
{Consider} the worst case{, in which} all the APs start {operating} and consuming their batteries simultaneously.
{If} only one redundant UAV is used to maintain all the AP positions, {constraint} (\ref{eq:constraint1}) is changed to
\begin{eqnarray}
n(2 T_f(d_i) +T_{ES}) &<& T_{AP}~~(\forall i), \label{eq:constraint_min}
\end{eqnarray}
\noindent
which {becomes} more difficult to {satisfy} as $N$ increases.

Now{,} we compare the two UAV-AP joint models in Table \ref{UAV-AP_models}.
JNT-CH simply follows {constraints} (\ref{eq:constraint1}) to (\ref{eq:constraint3}) and (\ref{eq:constraint_min}).
{In the rest of the paper,} we assume that the time {consumed} for battery replacement is {negligible} because it was reported that it took only {approximately} ten seconds {in} a prototyped system \cite{Toksoz2011} {and} will be shorter in a commercialized version.
{Therefore}, in JNT-RP, $T_{ES}$ {is} zero in {constraints} (\ref{eq:constraint1}) and (\ref{eq:constraint_min}).
JNT-RP can satisfy {constraint} (\ref{eq:constraint_min}) and achieve the minimum number of required UAVs, {namely,} $N$+1, {more easily} than JNT-CH, which will be examined later {through} simulation evaluation.

\subsubsection{Heuristic scheduling algorithm}\label{scheduling_joint}
Finding the optimal scheduling rule
of UAVs directly from the optimization problem (\ref{eq:objective}) is intractable {because} the battery discharging and charging functions are time-varying and nonlinear \cite{Traub2016}. Therefore, we focus on heuristic algorithms {--} one for JNT-CH and one for JNT-RP.
Every UAV in the system is autonomously operated with these algorithms in a distributed manner.
The algorithms {are used} mainly to determine how UAVs are associated with AP positions and when UAVs fly back to the ES.
Note that we assume {that the} information sharing necessary for scheduling UAVs in the system {is} ideally {performed} under the supervision {of} the BS or ES.

\begin{figure}[t]
\begin{center}
\includegraphics[width=\linewidth]{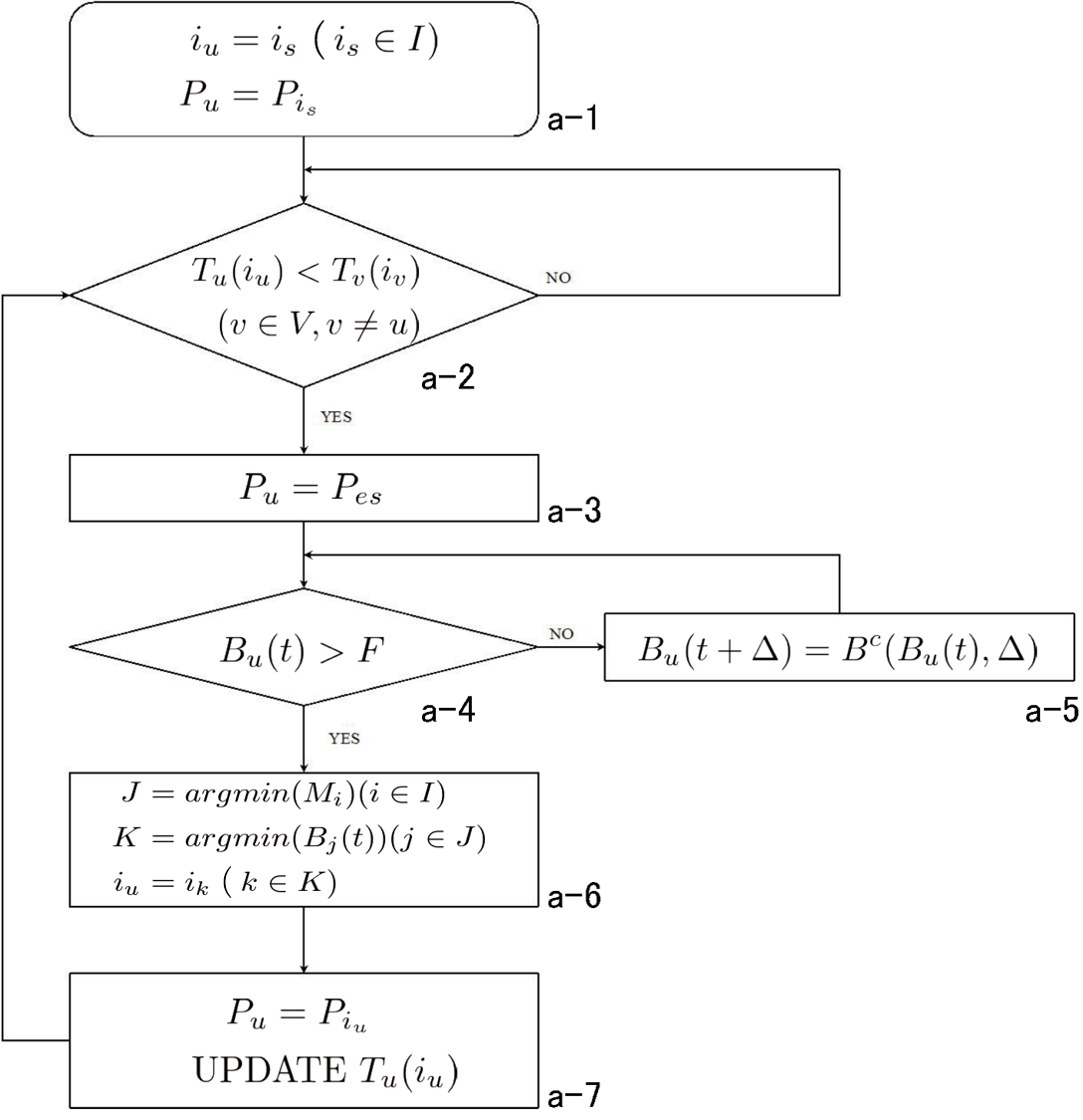}
\caption{Flowchart of {the} scheduling algorithm for JNT-CH}
\label{fig:flowchart_joint}
\end{center}
\end{figure}

Figure \ref{fig:flowchart_joint} shows the flowchart {for} the scheduling algorithm of UAVs for JNT-CH. 
The notations used in the figure are defined in Table \ref{flowchart_parameters}.
In the initial state{, which is denoted as} a-1, the operated UAV is associated with one of the predetermined AP positions, and the initial position of the operated UAV is the associated AP position.
The next step{, which is denoted as} a-2, judges whether another UAV $v$ has arrived at the associated AP position after the operated UAV {has} arrived.
If {so}, the operated UAV {leaves} for and {arrives} at the ES, as shown in a-3.  In a-4 and a-5, the operated UAV is charged {at the ES} until its battery becomes full. The next step{, namely, a-6,} determines {the} AP position {with which} the operated UAV should be associated. $J$ is the set of AP positions with which the number of associated UAVs is the smallest.
$K$ is the set of AP positions in $J$ {at which the} remaining  battery capacity is the smallest. 
{In Figure \ref{fig:flowchart_joint}}, $i_u=i_k$ ($k \in K$) means that the operated UAV is associated with one of the AP {positions} in $K$. 
After determining the associated AP position, the operated UAV leaves for and arrives at that position. $T_u(i_u)$ is updated to be used in a-2. Every UAV independently repeats {these} steps until it {receives a termination instruction}.
The scheduling algorithm for JNT-RP is similar to {the algorithm} for JNT-CH described above because both follow the UAV-AP joint model{, although} JNT-RP adopts battery replacement. 
The difference between {the algorithms for} JNT-RP and JNT-CH is that a-4 is replaced {with} $B_u(t)=B_{max}(t)$ {in the algorithm for JNT-RP}, 
which means {that} UAV $u$ replaces its battery {with} the {most-charged} one at the ES,
and a-5 is removed. 


\begin{table}[bt]
    \caption{Notations in flowcharts}
    \label{flowchart_parameters}
 \begin{tabular}{|l|p{5cm}|}
 \hline
{Parameter} & Value \\ \hline
$i=0,1,2,\cdot\cdot\cdot$ & {AP} position number \\ \hline
$I$ & Set of $i$ \\ \hline
$P_i=(X_i,Y_i,Z_i)$ & Position of $i$ \\ \hline
$M_i$ & Number of UAVs associated with $i$\\ \hline
$B_i(t)$ & Remaining battery capacity of device {(AP or UAV)} located {at} $i$ at time $t$\\ \hline
$T_v(i)$ & Latest time when UAV $v$ arrived at $i$ \\ \hline
$u$ & Operated UAV \\ \hline
$V$ & Set of UAVs\\ \hline
$i_u$ & $i$ associated with $u$\\ \hline
$P_u=(X_u,Y_u,Z_u)$ & Position of $u$\\ \hline
$B_u(t)$ & Remainig battery capacity of $u$ at time $t$\\ \hline
$B^c(B(t),\Delta)$ & Function of charging. Return $B(t)$ plus charging amount for $\Delta$ time\\ \hline $B^d(B(t),\Delta)$ & Function of discharging. Return $B(t)$ minus discharging amount for $\Delta$ time\\ \hline
$F$ & Full battery capacity\\ \hline
$P_{es}=(X_{es},Y_{es},Z_{es})$ & Position of ES\\ \hline

  \end{tabular} 
\end{table}

\subsection{UAV-AP separate case}\label{sec:separate}
\subsubsection{System model}\label{II-A}\label{sec:system_separate}
\begin{figure}[t]
 \begin{center}
  \includegraphics[width=0.9\linewidth]{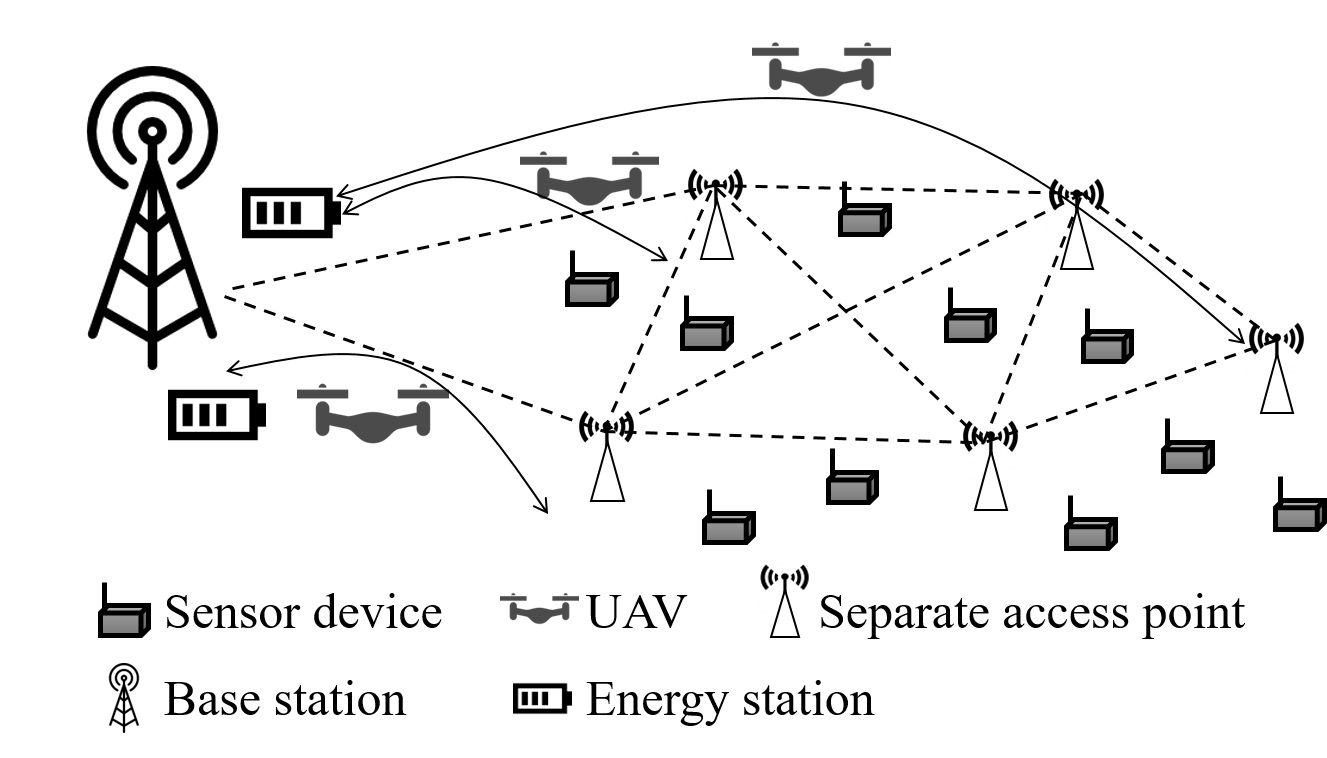}
  \caption{System model in {the} UAV-AP separate case}
  \label{fig:separate_model}
 \end{center}
\end{figure}

The proposed system in the UAV-AP separate case is illustrated in Figure~\ref{fig:separate_model}. 
The basic components of the system are {the same as} those in the joint case (Figure~\ref{fig:joint_model}).
In the separate case, UAVs carry and place separate APs {at} the predetermined AP positions in advance.
After that, UAVs are operated only {to maintain} the sustainability of the network {of pre-allocated} APs.

SPT-CH and SPT-RP in Table \ref{UAV-AP_models} follow the UAV-AP separate model, in which APs are initially placed at the predetermined positions and UAVs work {to maintain} AP and UAV batteries.
In SPT-CH, UAVs charge their batteries at the ES{, as in JNT-CH, and provide} the energy capacity for APs from their batteries.
In SPT-RP, UAVs are allowed to replace their batteries {with well-charged batteries} in advance at the ES {and swap} batteries with APs to sustain the lifetimes of APs.
Therefore, in SPT-RP, UAVs need to be equipped with a function that enables {them} to swap batteries with APs.
In comparison with the UAV-AP joint model, the UAV-AP separate model {benefits from a reduced} number of required UAVs; once separate APs are placed at the predetermined positions, only {the} UAVs necessary for maintaining AP batteries {are} required.

\subsubsection{Problem formulation}\label{sec:formulation_separate}
The optimization problem and the constraints in the UAV-AP separate case can be {represented by} (\ref{eq:objective}) and (\ref{eq:constraint1}) to (\ref{eq:constraint3}), respectively, {as in} the UAV-AP joint case.
However, {in the UAV-AP separate model,} since separate APs are {allocated} to the AP positions in advance, $M$ is {equal} to $m$.
Moreover, in the UAV-AP separate model, $T'_b$ in {constraint} (\ref{eq:constraint3}) is the remaining battery lifetime after the battery has been consumed {by} the AP.
{The baseline required number} of UAVs is $N$; $N$ UAVs are operated {to maintain} the batteries of $N$ separate APs.
Moreover, the minimum number of required UAVs is 1; only one UAV is used {to maintain} the sustainability of the network.
{Constraint} (\ref{eq:constraint_min}) is applicable as the constraint for {both} the UAV-AP separate case {and} the UAV-AP joint case.

{We} compare the two UAV-AP separate models in Table \ref{UAV-AP_models}.
In SPT-CH, an additional term, {namely,} $T_{UA}$, should be added to {constraints} (\ref{eq:constraint1}) and (\ref{eq:constraint_min}); $T_{UA}$
is the time {required to complete} the battery charge from the UAV to the AP.
Since we assume that the time {required} for battery replacement is {negligible}, in SPT-RP, $T_{ES}$ {is} zero in {constraints} (\ref{eq:constraint1}) and (\ref{eq:constraint_min}).
{Therefore,} SPT-RP can satisfy {constraint} (\ref{eq:constraint_min}) and achieve the minimum number of required UAVs, which is 1, more easily than SPT-CH.
{This will be} examined later thorough simulation evaluation.

\subsubsection{Heuristic scheduling algorithm}\label{scheduling_separate}
\begin{figure}[t]
\begin{center}
\includegraphics[width=\linewidth]{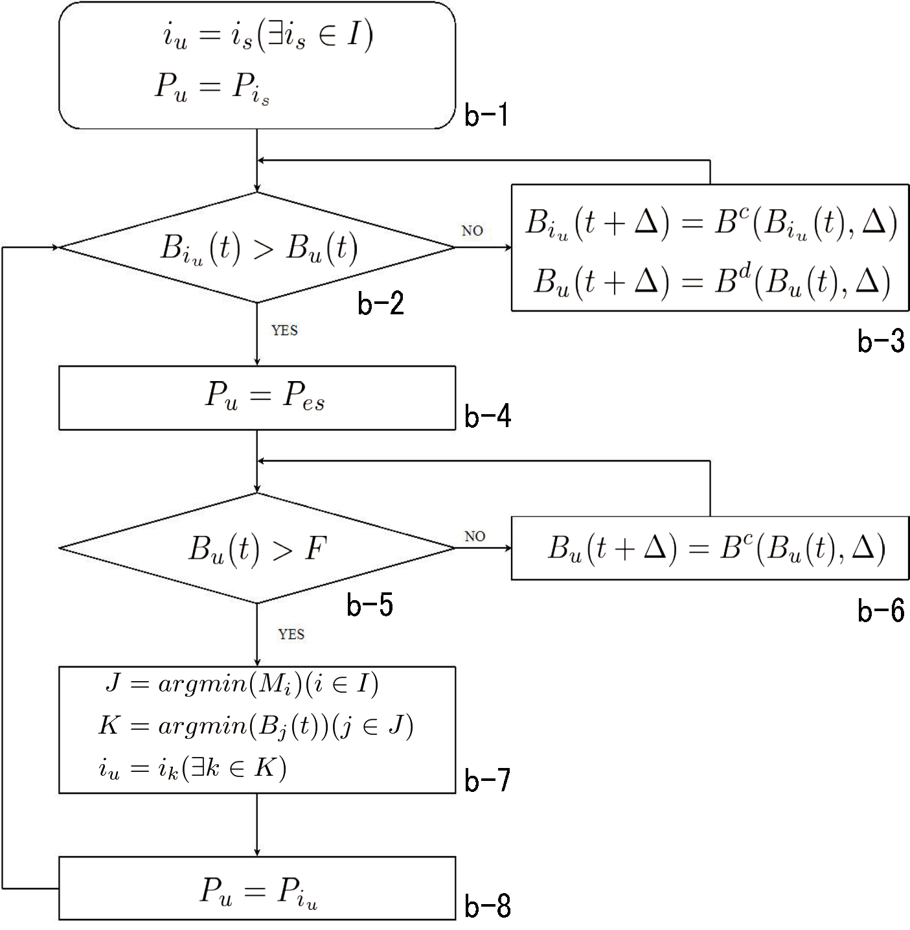}
\caption{Flowchart of {the} scheduling algorithm for SPT-CH}
\label{fig:flowchart_separate}
\end{center}
\end{figure}

Figure \ref{fig:flowchart_separate} shows {a} flowchart of SPT-CH, which is the UAV-AP separate and {battery-charged} model.
The notations listed in Table \ref{flowchart_parameters} are also used here.
The initial state{, which is denoted as b-1,} is the same as a-1. The next step{, which is denoted as b-2,} judges whether the remaining battery capacity of the associated AP is larger than that of the operated UAV. 
{If not so,} the operated UAV charges the associated AP in b-3. Although {the} other steps in SPT-CH are almost the same as those in JNT-CH, UPDATE $T_u(i_u)$ is not required because APs are separated from UAVs in SPT-CH. 
{In} the flowchart of SPT-RP, which is the UAV-AP separate and {battery-replacement} model, b-2 {is replaced with} $B_{i_u}(t)=B_u(t),B_u(t)=B_{i_u}(t)$, and b-3 {is} removed. In addition, b-5 {is replaced with}  $B_u(t)=B_{max}(t)$ {and} b-6 {is} removed in SPT-RP.


\section{Feasibility evaluation} 
\label{sec:feasibility}
%
This section {presents} and discusses the simulation model {for evaluating} the feasibility of the proposed system {and the results}.
We used the number of UAVs required {to} maintain the system, which was the objective in the problem formulation in the previous section, as the metric for the evaluation.
According to the problem formulation in the previous section, the required number of UAVs depends on the AP positions, the ES position, and the scheduling rule; it does not depend on the locations and the traffic characteristics of sensor devices, which will be considered in the throughput evaluation {in} Section \ref{sec:throughput}.
We examine {the models listed} in Table \ref{UAV-AP_models} and discussed in the previous section: JNT-CH, JNT-RP, SPT-CH, and SPT-RP.
UAVs are operated {in accordance with} the scheduling algorithms for each model{, which are} described in Sections \ref{scheduling_joint} and \ref{scheduling_separate}.

\subsection{Simulation model}\label{sec:sim_model1}
\begin{figure}[t]
 \begin{center}
  \includegraphics[width=\linewidth]{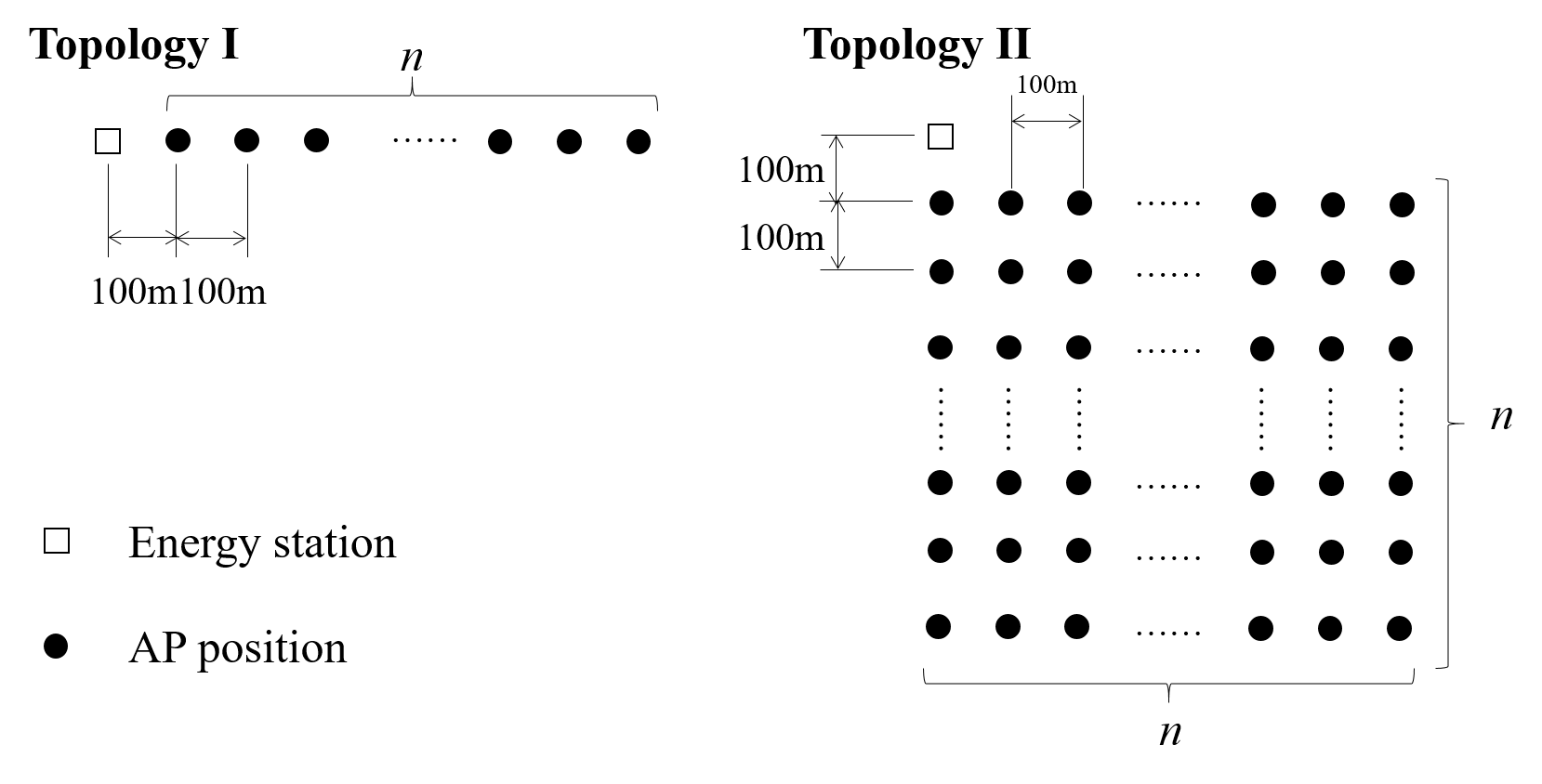}
  \caption{Simulation model for {the} feasibility evaluation}
  \label{simulation_model}
 \end{center}
\end{figure}

In our simulations, we used the two types of mesh topologies illustrated in Figure~\ref{simulation_model}. 
In topology I, {the} AP positions are placed in alignment at intervals of 100 m.
The number of AP positions $N$ is equal to $n$. 
In topology II, {the} AP positions are placed in {a} grid, where the intervals of both row and column placements are 100 m. 
Therefore, in topology II, the number of AP positions $N$ is equal to $n^2$.
Table \ref{battery_simulation parameters} summarizes the parameters {used} in our simulations, which we set not far from the specifications of recently commercialized UAVs \cite{bebop2}.
We assumed that {the} separate APs and UAVs working as APs in the joint model consume a constant {amount of} energy for {communicating}.
As we mentioned, we assumed that the time {consumption} for battery replacement is {negligible}.
Our simulator was developed using C++.

\begin{table}[bt]
    \caption{Simulation parameters for {the} feasibility evaluation}
    \label{battery_simulation parameters}
    \begin{center}
 \begin{tabular}{|l|l|}
 \hline
{Parameter} & Value \\ \hline
Battery capacity & 2700[mAh] \\ \hline
Battery power consumption when flying & 18[W] \\ \hline
Battery power consumption when commun. & 2[W] \\ \hline
Flying speed of UAV & 15[m/s] \\ \hline
Battery charging efficiency from UAV to AP & 100[\%] \\ \hline
Number of charging ports of ES & {Infinite} \\ \hline
  \end{tabular} 
  \end{center}
\end{table}

\begin{figure}[t]
 \begin{center}
  \includegraphics[width=.9\linewidth]{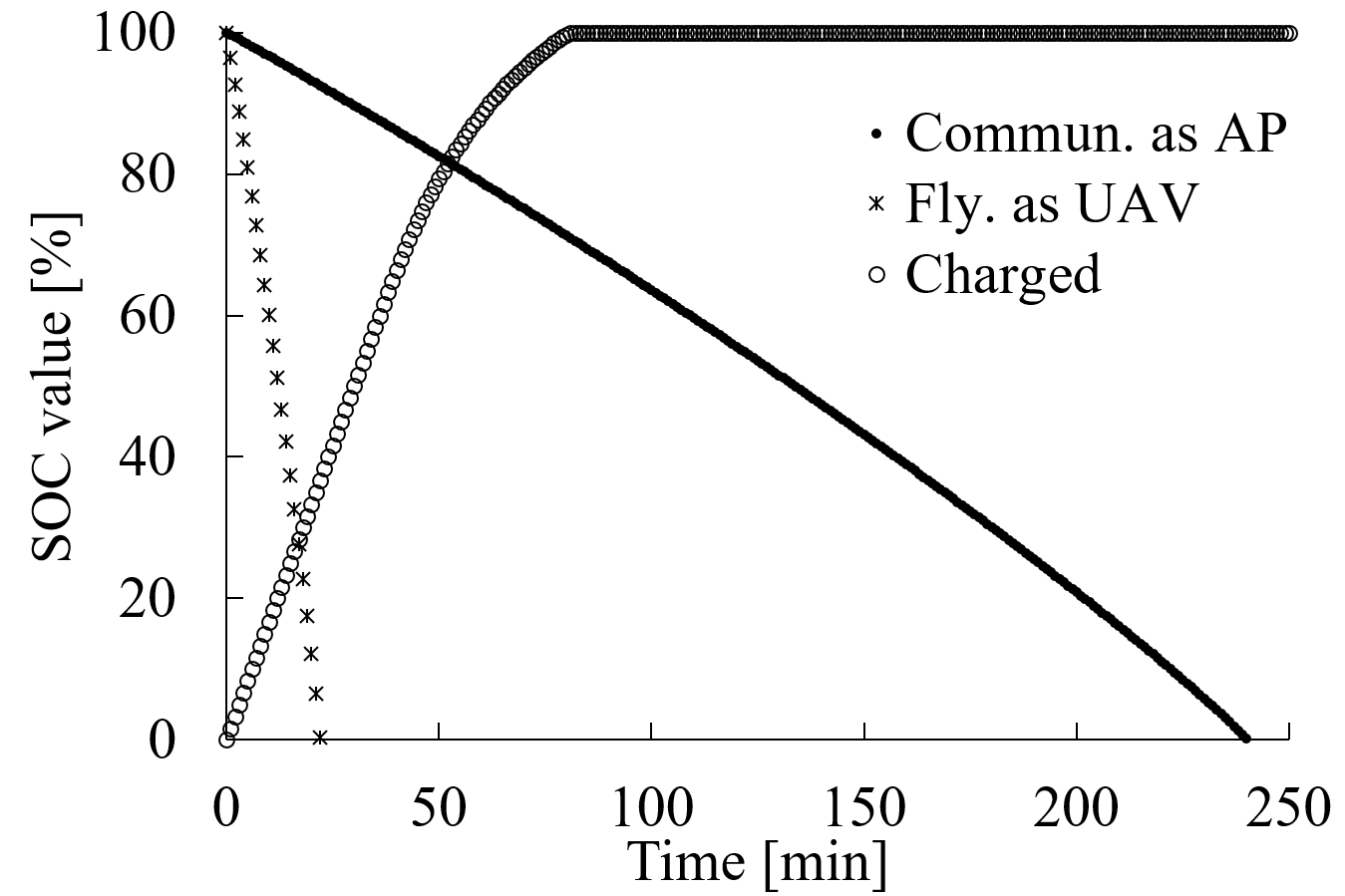}
  \caption{Charge and discharge characteristics of {the} battery model}
  \label{battery_model}
 \end{center}
\end{figure}

\subsection{Battery model}\label{V-A}
Given that {battery discharge models} applicable to our simulations are not available yet, we developed one{, which is} described next. 
Specifically, we developed a numerical model that enables us to reproduce the nonlinear discharge {characteristics} of realistic batteries.
%
The proposed discharge model is defined as a function {denoted as} \textit{discharge}($X_i$, $P$, $t_s$), where $X_i$, $P$, and $t_s$ are the {initial state-of-charge} (SOC) value in \%, the constant discharge {power}, and the {discharge duration}, respectively.
The output of \textit{discharge}($X_i$, $P$, $t_s$) is the SOC value of the battery that has been discharged with $P$ for $t_s$ after the initial state.
Traub reported that, although $D-V$ (discharge capacity -- voltage) curves are different for different current values $I$, $D-VI^n$ {forms a} unique curve upon selection of an appropriate {value of $n$,} regardless of current $I$ \cite{Traub2016}.
We assumed the use of a {lithium-ion} battery (Panasonic UPF614496; 2700 mAh capacity)
{and} used the real measurement result of the $D-V$ curve given in \cite{UPF614496}.
%
{For $n = 0.081$ ({which is} determined using the least-squares method),} the $D-VI^n$ curve can be approximated by
\begin{eqnarray}
VI^n  = & 3.623 \times 10^{-6}D^{2} - 2.191D + 3.643 \nonumber\\
& (0\leq D<200), \label{battery1} \\
VI^n  =  & \frac{3.614 - 0.325D + 1.114 \times 10^{-4}}{1-0.094D + 1.97 \times 10^{-5} + 3.964 \times 10^{-9}}   \nonumber\\
& (200\leq D) \label{battery2}
\end{eqnarray}
{By equations (\ref{battery1}) and (\ref{battery2}), we can calculate} the remaining capacity of a battery at {any time,} even after the battery {has been} consumed {at} different power levels for different purposes, i.e., flying and communicating.
In Figure~\ref{battery_model}, the characteristics of the discharged battery in the cases of flying and communicating are plotted.

We also {developed} a numerical battery charge model, {which} enables us to reproduce the nonlinear charge characteristic of realistic batteries.
The proposed model is represented as a function {denoted as} \textit{charge}($X_i$, $t_{ref}$), where $X_i$ {(in \%)} and $t_{ref}$ {(in seconds)} are the SOC value and {charge duration}, respectively.
The output of \textit{charge}($X_i$, $t_{ref}$) is the SOC value of {a} battery that has been charged for $t_{ref}$ after the initial state.
%
For validating the numerical model, we used the real $t-C$ (time -- charge capacity) curve of the {lithium-ion} battery given in \cite{UPF614496}.
If the battery is charged {using the constant-voltage/constant-current method,} the $t-C$ curve can be approximated by
\begin{eqnarray}
C  = & \frac{2700}{3600}t~~~~(0<t<2238),  \label{battery3} \\
C  = & \frac{2700}{3600} \times 2238 +  \nonumber \\
& \frac{2690}{60} \frac{1}{0.0336} \{1-\exp(-0.0336 \times 60t)\} \nonumber \\
& (2238\leq t) \label{battery4}
\end{eqnarray}
From these equations, we obtain the remaining capacity of a battery {that has been} charged for a {specified amount of time} after the initial {state} of SOC.
In Figure~\ref{battery_model}, the characteristic {curve} of the charged battery is plotted.

\begin{figure}[t]
 \begin{center}
  \includegraphics[width=.9\linewidth]{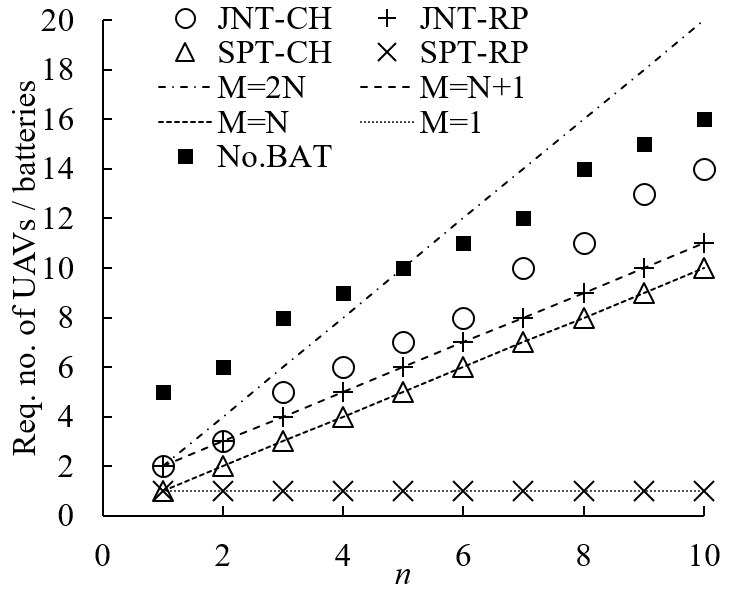}
  \caption{{Numbers} of required UAVs and batteries in topology I}
  \label{battery_tp1}
 \end{center}
\end{figure}

\begin{figure}[t]
 \begin{center}
  \includegraphics[width=.9\linewidth]{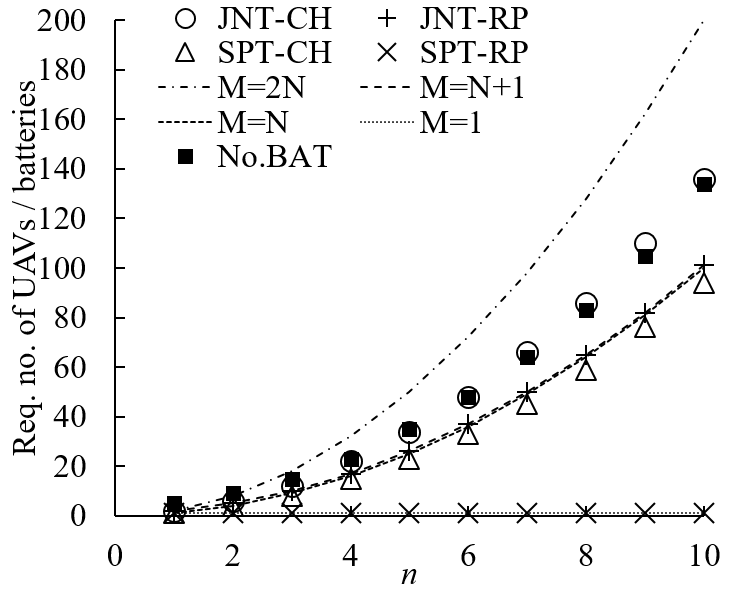}
  \caption{{Numbers} of required UAVs and batteries in topology II}
  \label{battery_tp3}
 \end{center}
\end{figure}

\subsection{Results}\label{IV-B}
Figure \ref{battery_tp1} plots the number of required UAVs in the four models, {namely}, JNT-CH, JNT-RP, SPT-CH, and SPT-RP, {for} topology I.
The horizontal axis {corresponds to} $n$ in Figure~\ref{simulation_model}.
First, let us observe the results {for} the two UAV-AP joint models: JNT-CH and JNT-RP. 
As we {explained} in Sections \ref{sec:formulation_joint} and \ref{sec:formulation_separate}, the theoretical baseline and {lower bound} for these models are $M=2N$ and $M=N+1$, respectively.
Compared with these two theoretical benchmarks, JNT-CH {requires} a smaller number of UAVs than the baseline, while JNT-RP {achieves the lower bound}.
{As we discussed in the previous section,} since $T_{ES}$ in JNT-RP is zero, it can easily satisfy {constraint} (\ref{eq:constraint_min}), which is why the number of required UAVs {is} reduced to the {lower bound}.
On the other hand, to satisfy the requirement given by {constraint} (\ref{eq:constraint1}), JNT-CH needs a {specific} number of UAVs between the baseline and the {lower bound}.
{Next,} we consider the results of the two UAV-AP separate models: SPT-CH and SPT-RP.
The baseline and the theoretical {lower bound} for the UAV-AP separate models are $M=N$ and $M=1$, respectively.
As shown in Figure~\ref{battery_tp1}, the number of required UAVs in SPT-CH is almost {the} same as {in} the baseline, which means {that} SPT-CH did not work efficiently.
This is because, as we mentioned in Section \ref{sec:formulation_separate}, SPT-CH {incurs} additional overhead 
because of the charging time from UAVs to APs, {namely,} $T_{UA}$, which is not included in SPT-RP.
{In contrast,} SPT-RP achieves the {lower bound}.
Finally, we discuss the number of required batteries in JNT-RP and SPT-RP, which is also plotted in Figure~\ref{battery_tp1}.
It was {surprising} that the number of required batteries {was the same for both}.
When a new UAV arrives at an AP position, in JNT-RP, {the UAV working as an AP at that position leaves for the ES, carrying its discharged battery.}
{In contrast, in SPT-RP, separate APs are preallocated at the positions, and the discharged batteries are carried by UAVs, which is essentially the same as in JNT-RP.}
In both JNT-RP and SPT-RP, at least, $n$ batteries are necessary for $n$ AP positions.
Therefore, if we subtract $n$ from the number of required batteries plotted in Figure~\ref{battery_tp1}, we {observe} that only 3 to 5 {additional} batteries are required.

Figure \ref{battery_tp3} plots the {results} for topology II.
{The trends are {almost} the same as those shown in Figure~\ref{battery_tp1}:}
1) {the} required numbers of UAVs in both the UAV-AP joint and separate models were between the {lower bound} and the baseline for each model; 
2) JNT-RP and SPT-RP achieved the {lower bounds}; {and}
3) SPT-CH did not reduce the number of required UAVs compared with the baseline because of its battery charging overhead from UAVs to APs.
However, unlike {in} Figure~\ref{battery_tp1}, quadric-like increases in the numbers of required UAVs and batteries vs. $n$ {are shown} in Figure~\ref{battery_tp3}.
This is {because} the number of AP positions $N$ increases proportionally {with} $n^2$.
The additional number of batteries required in JNT-RP and SPT-RP in topology II {ranged from} 3 to 33, which is much larger than in the case of topology I because the number of AP positions is much larger.

In summary, JNT-RP and SPT-RP require {only} the lower-bound number of UAVs.
They can be operated with only 3 to 5 additional batteries in topology I, while 3 to 33 additional batteries are necessary to sustain topology II.
However, as we mentioned in Section \ref{sec:system_separate}, a technical difficulty {is encountered} in SPT-RP: UAVs need to be equipped with a function that enables {them} to swap batteries with APs, which is not necessary in the other three models.
Therefore, {we conclude} that JNT-RP {is} the best option.

\section{Throughput evaluation}
\label{sec:throughput}

\subsection{Simulation model}\label{VI-A}
In the previous sections, we used the number of required UAVs and the number of required batteries as {metrics} to examine the feasibility of our system. 
In this section, using throughput as the metric, we discuss how our system works as a wireless mesh network that delivers sensing data.
%
%
We used QualNet, which is a {well-known} off-the-shelf software for wireless network simulations \cite{Qualnet}, to measure throughput.
We assumed a scenario {in which} distributed sensor devices upload their sensor data to {a} BS located at the same position {as} the ES in Figure \ref{simulation_model} via the nearest AP {to} each sensor device. 
However, we only {observed} throughputs of aggregated data from each AP to the BS, which we {thought} {was sufficient for evaluating} the capability of the network {in} collecting sensor data because we can assume {that} transmissions from sensor devices to APs {do} not affect the aggregated throughputs {if} the frequency channel assigned to the sensor-AP transmission is isolated from AP-AP and AP-BS transmissions.
We measured the average throughput from each AP to the BS and the total throughput to {determine} an appropriate range of $n$ in topologies I and II in terms of the capability {of} collecting sensor data.
\par

{The} simulation parameters are listed in Table~\ref{throughput simulation parameters}.
{We} adopted IEEE802.11g as the wireless interface for AP-AP and AP-BS transmissions {because} its transmission rate is determined {on the basis of} the channel model more simply than more recent specifications {such as} 11n and 11ac \cite{Li2013}, which use the {multiple-input and multiple-output} (MIMO) technology \cite{Jones2015}.
The ad hoc mode of IEEE802.11g is used for AP-AP and AP-BS links.
\par
As the network-level setting, the {ad hoc on-demand distance vector} (AODV) \cite{AODV} is used as the routing protocol in the wireless mesh networks.
AODV is a reactive routing protocol; it does not maintain routing tables continuously but produces them only when data {must} be forwarded, which we {believe} is suitable for our system because the topology and nodes of the network can dynamically change.
\par
As the application-level setting, to simulate traffic flows of aggregated sensor data from APs to the BS, constant-bitrate (CBR) traffic was generated at each AP.
To keep {the total traffic load} in the network constant, we set the bitrate of the CBR traffic to 24 Mbps divided by the number of APs, where 24 Mbps is close to the effective transmission rate of 11g {with} 54 Mbps as the physical transmission rate.

\begin{table}[b]
    \caption{Simulation parameters for throughput evaluation}
    \label{throughput simulation parameters}
    \begin{center}
 \begin{tabular}{|l|l|}
 \hline
{Parameter} & Value \\ \hline
Wireless standard & IEEE802.11g \\ \hline
No. of {channels} & 1 \\ \hline
Frequency & 2.4[GHz] \\ \hline
Physical transmission rate & 54[Mbps] \\ \hline
Pathloss model & Two-ray \\ \hline
Routing protocol & AODV \\ \hline
Rate of CBR & 24Mbps / no. of APs \\ \hline
  \end{tabular} 
  \end{center}
\end{table}

\subsection{Results}\label{VI-B}
\begin{figure}[t]
 \begin{center}
  \includegraphics[width=0.8\linewidth]{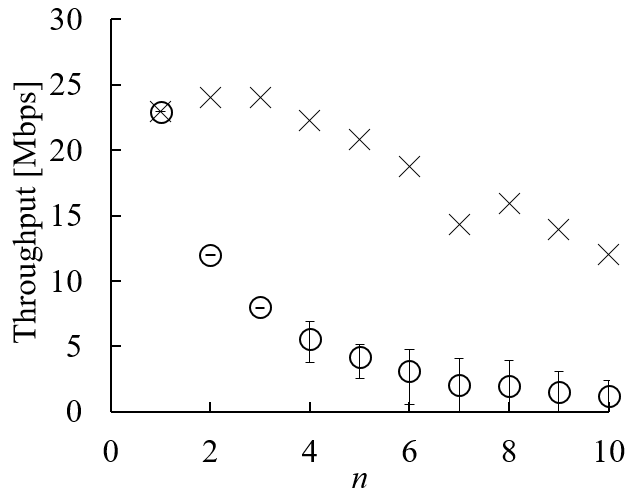}
  \caption{Throughput performance in topology I. Total throughput, average throughput, and maximum/minimum throughputs are plotted using {$\times$s, circles,} and error bars{, respectively}.}
  \label{throughput_tp1_1}
 \end{center}
\end{figure}

Figure \ref{throughput_tp1_1} {plots throughput} versus $n$ in topology I. 
%
%
{The} total throughput was high in the range of $n \leq 4${,} while it decreased as $n$ increased when $n \geq 5$.
This is because as $n$ increases, channel contention and frame collision occur more easily, which results in decreased throughput.
{The} average throughput degraded drastically in the figure as $n$ increased.
{According to the error bars,} the difference between the maximum and minimum throughputs was small when $n \leq 3$ {and increased} when $n \geq 4$.
{The} minimum throughput was zero when $n \geq 7$, which {indicates that} some APs could not upload their data to the BS.
Since the maximum number of hopcounts from APs to the BS increases as $n$ increases, it {is} more difficult for APs further from the BS to deliver their data to their destination ({the} BS).
{This} suggests that, in {practice}, we should operate our system in the range of $n \leq 6$.
{To} operate the network in the range of $n \geq 7$, we should {limit the} uploading rates from APs so that APs far from the BS can deliver their data to the BS.

\par
\begin{figure}[t]
\begin{center}
\subfigure[$1\leq n\leq10$]{\includegraphics[width=0.8\linewidth]{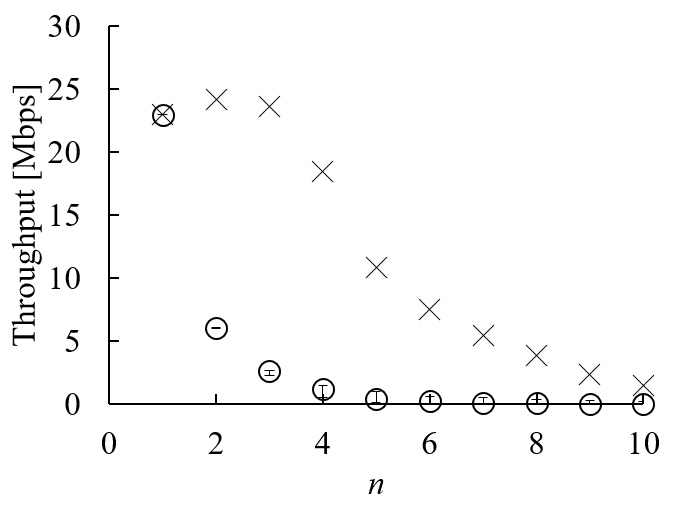}}
\subfigure[$n \geq 5$]{\includegraphics[width=0.8\linewidth]{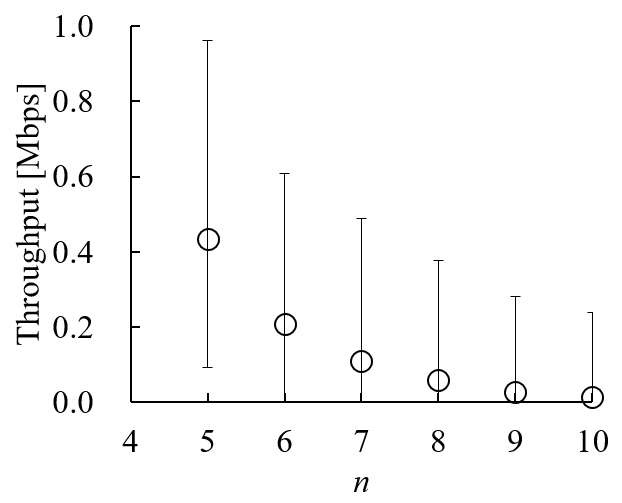}}
\caption{Throughput performance in topology II. Total throughput, average throughput, and maximum/minimum throughputs are plotted using {$\times$s, circles,} and error bars, {respectively}.}
\label{throughput_tp3_2}
\end{center}
\end{figure}

Figure \ref{throughput_tp3_2} (a) {plots throughput} versus $n$ {in topology II}.
Note that in topology II, the number of APs $N$ is equal to $n^2$.
%
%
{High} total throughput was available when $n\leq3${,} while it degraded as $n$ increased when $n\geq4$.
This is {because} as $n$ becomes large, the network becomes more congested.
To focus on the throughput performance in the case of large numbers of APs, we show the average/minimum/maximum throughputs in the range of $5\leq n\leq10$ in Figure~\ref{throughput_tp3_2} (b).
{The} average throughput was smaller than 1 Mbps.
{According to the error bars, the differences} in throughput among APs {were} large {when} $n\geq 5$.
{In particular}, when $n\geq 6$, the minimum throughput was zero, which {indicates that} some APs could not upload data to the BS.
Thus, in topology II, in {practice}, we should operate our system in the range of $n \leq 5$.
As we mentioned {for} Figure~\ref{throughput_tp1_1}, to operate the network in the range of $n \geq 5$, we should limit {the} uploading rates from APs so that APs far from the BS can deliver their sensor data to the BS.

\section{Conclusions}\label{VII}
This paper proposed a new design of wireless mesh networks formed by UAVs under the assumption that batteries and APs are replaceable and separable from UAVs and both are carried and placed {at appropriate} positions by the mechanical automation of UAVs.
We first {presented} possible models of UAV-formed multihop networks and {compared} them.
{Then, we} presented the problem formulation and the scheduling algorithm of UAVs.
Through computer simulations, we numerically evaluated {the number of UAVs that} each model requires {for maintaining} a wireless mesh network.
The simulation results suggested {that} the number of required UAVs is affected by the number of AP positions {and} can be minimized by introducing {a} battery replacement function. 
We also considered a realistic battery model and the required number of batteries and showed that our system {performs} well. 
Furthermore, throughput analysis demonstrated how our system {performs} as a wireless mesh-network infrastructure for sensor nodes.
Future work {will include} implementation and experimental evaluation of the proposed system.


\section*{Acknowledgement}
This work was partly supported by 
JSPS KAKENHI Grant No. JP25730057 and No. 17H01732.\\
The authors would like to thank Mr. Yuki Goto, who graduated from Kyoto University in March 2017, and Ms. Miwa Tobita, who is currently a student of Kyoto University, for their suggestions.


\medskip
\includegraphics[width=1in,height=1.25in,clip,keepaspectratio]{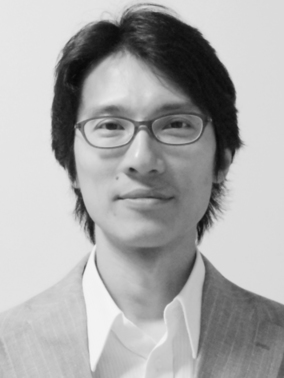}
\noindent {\bf Ryoichi Shinkuma}
received the B.E., M.E., and Ph.D. degrees in Communications Engineering from Osaka University, Japan, in 2000, 2001, and 2003, respectively.
In 2003, he joined the faculty of Communications and Computer Engineering, Graduate School of Informatics, Kyoto University, Japan, where he is currently an Associate Professor.
He was a Visiting Scholar at Wireless Information Network Laboratory (WINLAB), Rutgers, the State University of New Jersey, USA, from 2008 Fall to 2009 Fall.
His research interests include network design and control criteria, particularly inspired by economic and social aspects.
He received the Young Researchers' Award from IEICE in 2006 and the Young Scientist Award from Ericsson Japan in 2007, respectively.
He also received the TELECOM System Technology Award from the Telecommunications Advancement Foundation in 2016.
He has been the chairperson of the Mobile Network and Applications (MoNA) Technical Committee of IEICE Communications Society since June 2017.
He is a member of IEEE.


\medskip
\includegraphics[width=1in,height=1.25in,clip,keepaspectratio]{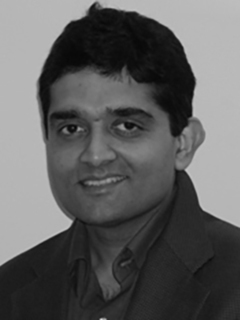}
\noindent{\bf Narayan B. Mandayam} (S'89-M'94-SM'99-F'09) received the B.Tech (Hons.) degree in 1989 from the Indian Institute of Technology, Kharagpur, and the M.S. and Ph.D. degrees in 1991 and 1994 from Rice University, all in electrical engineering.  Since 1994 he has been at Rutgers University where he is currently a Distinguished Professor and Chair of the Electrical and Computer Engineering department. He also serves as Associate Director at WINLAB. He was a visiting faculty fellow in the Department of Electrical Engineering, Princeton University, in 2002 and a visiting faculty at the Indian Institute of Science, Bangalore, India in 2003. Using constructs from game theory, communications and networking, his work has focused on system modeling, information processing and resource management for enabling cognitive wireless technologies to support various applications. He has been working recently on the use of prospect theory in understanding the psychophysics of pricing for wireless data networks as well as the smart grid.  His recent interests also include privacy in IoT, resilient smart cities as well as modeling and analysis of trustworthy knowledge creation on the internet.
Dr. Mandayam is a co-recipient of the 2015 IEEE Communications Society Advances in Communications Award for his seminal work on power control and pricing, the 2014 IEEE Donald G. Fink Award for his IEEE Proceedings paper titled ``Frontiers of Wireless and Mobile Communications'' and the 2009 Fred W. Ellersick Prize from the IEEE Communications Society for his work on dynamic spectrum access models and spectrum policy. He is also a recipient of the Peter D. Cherasia Faculty Scholar Award from Rutgers University (2010), the National Science Foundation CAREER Award (1998) and the Institute Silver Medal from the Indian Institute of Technology (1989). He is a coauthor of the books: Principles of Cognitive Radio (Cambridge University Press, 2012) and Wireless Networks: Multiuser Detection in Cross-Layer Design (Springer, 2004). He has served as an Editor for the journals IEEE Communication Letters and IEEE Transactions on Wireless Communications. He has also served as a guest editor of the IEEE JSAC Special Issues on Adaptive, Spectrum Agile and Cognitive Radio Networks (2007) and Game Theory in Communication Systems (2008). He is a Fellow and Distinguished Lecturer of the IEEE.

\end{document}